\documentclass[english,aps,prl,reprint,superscriptaddress,showpacs,amsfonts,amsmath,amssymb,floatfix]{revtex4-1}

\usepackage[latin9]{inputenc}
\usepackage{color}
\usepackage{babel}
\usepackage{units}
\usepackage{amsmath}
\usepackage{graphicx}
\usepackage[unicode=true,
 bookmarks=false,
 breaklinks=false,pdfborder={0 0 1},backref=false,colorlinks=true]
 {hyperref}
\hypersetup{
 citecolor=myblue,linkcolor=myred}

\makeatletter

\DeclareTextSymbolDefault{\textquotedbl}{T1}

\usepackage{dcolumn}
\usepackage{bm}
\usepackage{color}

\usepackage{psfrag,graphicx} 

\usepackage{epsf}
\usepackage{epstopdf}

\usepackage{appendix}
\usepackage{changes}

\usepackage[caption=false]{subfig}

\usepackage{xr}
\graphicspath{{figs/}}

\definecolor{mygrey}{gray}{0.35}
\definecolor{myblue}{rgb}{0.2,0.2,0.8}
\definecolor{myblue2}{rgb}{0,.447,.741}
\definecolor{myzard}{cmyk}{0,0,0.05,0}
\definecolor{mywhite}{rgb}{1,1,1}
\definecolor{myred}{rgb}{1,0.,0.3}
\definecolor{myred2}{rgb}{.85,.325,.098}
\definecolor{mygreen}{rgb}{.466,.674,.188}
\definecolor{mypurple}{rgb}{.494,.184,.556}

\def\be{\begin{equation}}
\def\ee{\end{equation}}
\def\ba{\begin{align}}
\def\enda{\end{align}}
\def\bi{\begin{itemize}}
\def\ei{\end{itemize}}

 \def\ee{\mathord{\rm e}}

 \def\ee{\mathord{\rm e}}

\renewcommand{\ee}{{\rm e}}

\def\beq{\begin{equation}}
\def\beq{\begin{equation}}
\def\eeq{\end{equation}}

\@ifundefined{showcaptionsetup}{}{%
 \PassOptionsToPackage{caption=false}{subfig}}
\usepackage{subfig}
\makeatother

\begin{document}
\global\long\def\bla#1{\left(#1\right)}
\global\long\def\blb#1{\left[#1\right]}
\global\long\def\blc#1{\left\langle #1\right\rangle }
\global\long\def\ket#1{\left|#1\right\rangle }
\global\long\def\bra#1{\left\langle #1\right|}
\global\long\def\braket#1#2{\left\langle #1|#2\right\rangle }
\global\long\def\sinc{\text{sinc}}

\title{Limits on spectral resolution measurements by quantum probes}

\author{Amit Rotem}

\thanks{These authors contributed equally.}

\affiliation{Racah Institute of Physics, The Hebrew University of Jerusalem, Jerusalem
91904, Givat Ram, Israel}

\author{Tuvia Gefen}

\thanks{These authors contributed equally.}

\affiliation{Racah Institute of Physics, The Hebrew University of Jerusalem, Jerusalem
91904, Givat Ram, Israel}

\author{Santiago Oviedo-Casado}

\affiliation{Departamento de F\'isica Aplicada, Universidad Polit\'ecnica de Cartagena,
Cartagena 30202, Spain}

\author{Javier Prior}

\affiliation{Departamento de F\'isica Aplicada, Universidad Polit\'ecnica de Cartagena,
Cartagena 30202, Spain}

\affiliation{Institute Carlos I for Theoretical and Computational Physics, Universidad
de Granada, Granada 18071, Spain}

\author{Simon Schmitt}

\affiliation{Institute for Quantum Optics, Ulm University, Albert-Einstein-Allee
11, Ulm 89081, Germany}

\author{Yoram Burak}

\affiliation{Racah Institute of Physics, The Hebrew University of Jerusalem, Jerusalem
91904, Givat Ram, Israel}

\affiliation{Edmond and Lily Safra Center for Brain Sciences, The Hebrew University
of Jerusalem, Jerusalem 91904, Givat Ram, Israel}

\author{Liam McGuiness}

\affiliation{Institute for Quantum Optics, Ulm University, Albert-Einstein-Allee
11, Ulm 89081, Germany}

\author{Fedor Jelezko}

\affiliation{Institute for Quantum Optics, Ulm University, Albert-Einstein-Allee
11, Ulm 89081, Germany}

\author{Alex Retzker}

\affiliation{Racah Institute of Physics, The Hebrew University of Jerusalem, Jerusalem
91904, Givat Ram, Israel}

\date{\today}
\begin{abstract}
The limits of frequency resolution in nano NMR experiments have been
discussed extensively in recent years. It is believed that there is
a crucial difference between the ability to resolve a few frequencies
and the precision of estimating a single one. Whereas the efficiency
of single frequency estimation gradually increases with the square
root of the number of measurements, the ability to resolve two frequencies
is limited by the specific time scale of the signal and cannot be compensated
for by extra measurements. Here we show theoretically and demonstrate
experimentally that the relationship between these quantities is more
subtle and both are only limited by the Cram\'er-Rao bound of a single
frequency estimation.
\end{abstract}
\maketitle
\begin{figure}
\subfloat[]{\includegraphics[width=0.46\columnwidth]{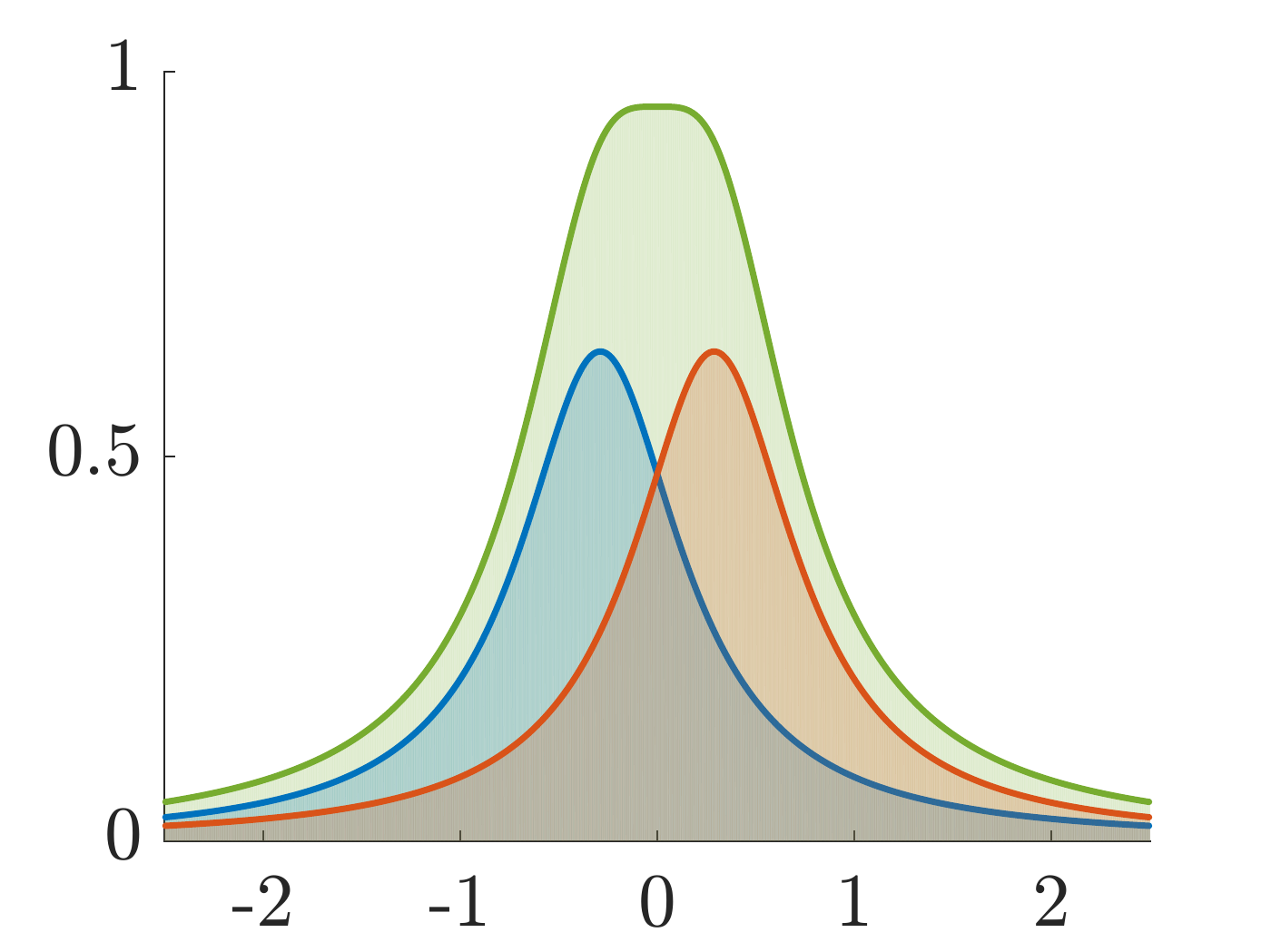} \label{ResLim_Rayleigh}}
\hfill
\subfloat[]{\includegraphics[width=0.46\columnwidth]{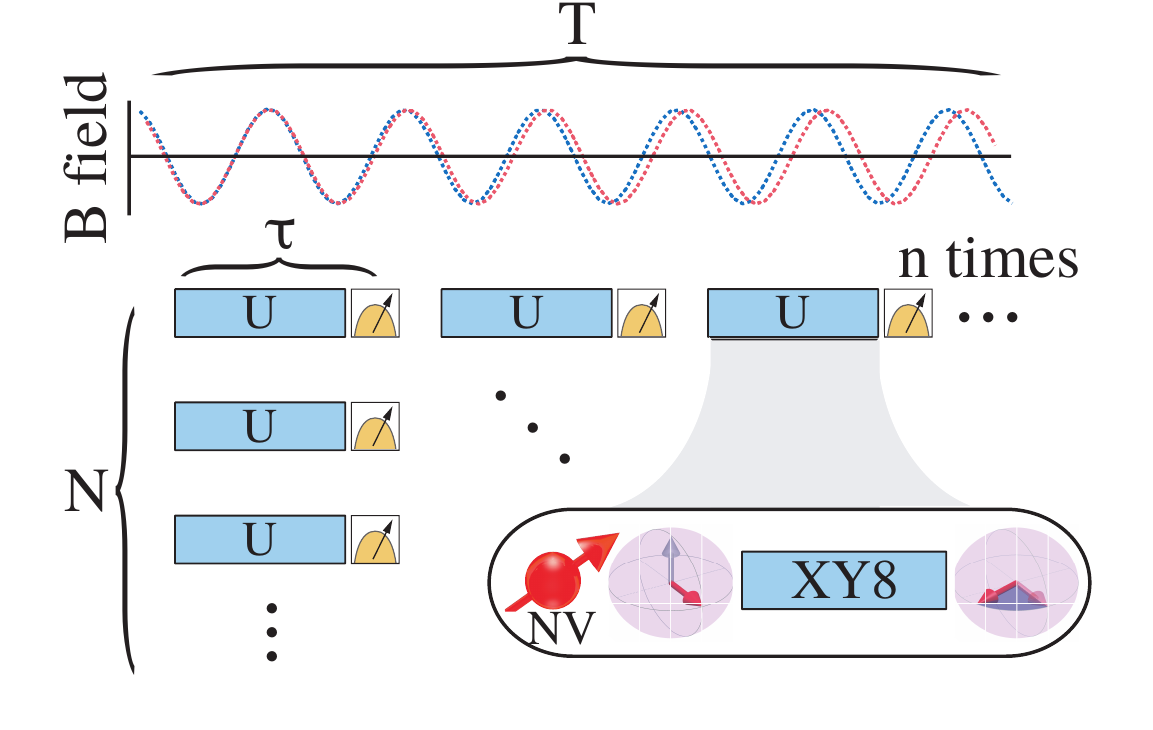} \label{scheme_block}}

\caption{\emph{Resolution problem.} (a) When the two signals are close enough
(blue and orange), their sum looks like one Lorentzian (green) making
it impossible to differentiate between two sources and one broad source.
\textit{Experimental scheme.} (b) Experimental scheme for a set of
frequencies. A set of $n$ measurements is made during the time that the
phase of the signal is stable, where the time of each measurement
is $\tau$ and is limited by the coherence time of the probe ($T_{2}$).
The sets are repeated $N$ times to obtain more statistics; i.e., $N$
uncorrelated sets. The estimation of the frequencies is done by calculating
the likelihood function and locating its maximum value \citep{cramer,kruschke2014doing,cover2012elements};
i.e., $\max\mathcal{L}(\{\delta_{k},\Omega_{k},\varphi_{k}^{\protect\bla{\ell}}\}_{k,\ell})$,
where $k$ is the index of the frequencies $\bla{\delta}$ and $\ell$ designates
the phases $\bla{\varphi}$, and $\Omega$ is the Rabi-frequency of the signal.}

\label{ResLim}
\end{figure}

We consider the problem of spectral resolution; i.e., differentiating
between two close frequency components of a signal. In the nano NMR
setting this can be formulated as follows. A time dependent signal
is coupled to a two-level system by a term such as $\mathcal{H}=f(t)\sigma_{z},$
where $\sigma_{z}$ is a Pauli matrix, with the aim to assess the spectral
content of $f(t).$ This problem has been extensively examined in
the past few years via NV centers in diamond \citep{Qdyne,boss2017,Pfender2016,Laraoui2013,ajoy2015,Bucher2017,Staudacher2013,muller2014,Aslam,Lovchinsky,Mamin}.
The limit of resolution of the frequency spectrum of signals is believed
to be set by the line-width of the power spectrum \citep{Qdyne,boss2017,Bucher2017} where the liquid state is dominated by diffusion \cite{pham,Staudacher2013,Aslam,Kong2015} . 
The main problem is illustrated in Fig.~\ref{ResLim_Rayleigh},
where two signals that are close enough manifest a power spectrum
which is similar to that of a single broad frequency. This intuition
that resolution is limited by the line-width is based on the Rayleigh
criterion from optics \citep{Abbe,Rayleigh} where an analogy is drawn
between the wavelength and the line-width. This notion is one of the
main pillars of spectroscopy. Here, we challenge this concept. 

The traditional method of spectroscopy with quantum sensors uses dynamical
decoupling pulses for a certain duration, since the fluorescence as a function
of the dynamical decoupling frequency reflects the spectrum of the
signal \citep{Kotler2011,Maze}. Thus when implementing this method
two frequencies are only resolvable if the difference between them
is larger than, roughly, $T_{2}^{-1},$ where $T_{2}$ is the coherence
time of the probe. This was believed to impose a fundamental limit
on frequency resolution. However, it was realized that by transferring
the quantum phase of the sensor to the state population that survives
up to longer $T_{1}$ relaxation times \citep{Staudacher2015,Kong2015},
resolution could be improved. Moreover, by using a hybrid quantum
system where an additional long-lived qubit acts as a more stable
clock \citep{Pfender2016,Laraoui2013,Unden2016,Zaiser2016,Rosskopf2016,Aslam}
the limit could be extended to the coherence time of the ancilla qubit.
Recently it was realized that the quantum memory could be replaced
by a classical one \citep{Qdyne,boss2017,Bucher2017,Khodas}. In these
contributions it was shown that although the efficiency of estimating
a single frequency improves with the number of measurements, the resolution
limit is set by the specific time scale of the scheme.

Here we show that by utilizing a suitable processing algorithm of the data, the behavior of resolution in the nano NMR setting is highly
similar to that of a single frequency estimation; i.e., it improves with
extra uncorrelated measurements and does not diverge for close frequencies. In this letter, we focus on the post-processing of the data, given the measurement 
protocol of \citep{Qdyne,boss2017,Bucher2017}.

\paragraph{Formulation of the problem and outline of the algorithm---}

\label{sec:formulation}

We define resolution as the ability to determine the number of frequencies
in a signal and estimating them. We consider in detail the case
in which the maximal number of frequencies is two i.e., whether the signal contains a single frequency or two different
frequencies, and test the extension to multiple frequencies numerically. The basic algorithm proceeds as follows. We first assume
the number of frequencies is two (denoted as $\delta_{1},\,\delta_{2}$)
and then use Maximum-likelihood (ML) \citep{ML} to either estimate
$\delta_{1}-\delta_{2}$ or $\delta_{1},\,\delta_{2}$ separately.
If $\Delta\left(\delta_{1}-\delta_{2}\right)<|\delta_{1}-\delta_{2}|$
or $\Delta\bla{\delta_{1}}+\Delta\bla{\delta_{2}}<|\delta_{1}-\delta_{2}|$,
where $\Delta\bla x$  is the standard deviation (SD) of $x$, then we can deduce that there are two
resolvable frequencies (as the error probability is negligible). Therefore
the quantities $\Delta\left(\delta_{1}-\delta_{2}\right),\:\Delta\bla{\delta_{1}},\:\Delta\bla{\delta_{2}}$
determine our resolution limits. On the other hand, precision is our
ability to estimate the value of a frequency given that the signal
contains only one frequency.

To address this problem of frequency resolution it is instructive
to first understand the origin of the notion that resolution and single
frequency estimation are two different things. The standard frequency
measurement protocol is based on dynamical decoupling \citep{freeman1998,slichter2013}.
In the regime where the coherence time of the signal is longer than
the coherence time of the probe, synchronized measurements can be performed
\citep{Qdyne,boss2017,Bucher2017}. In these scenarios the standard
estimation method involves estimating the frequencies out of the power
spectrum. For a single frequency $\delta$, the spectrum is a
peaked function $g\left(\left(x-\delta\right)T_{\varphi}\right)$
(e.g. $g\left(\cdot\right)$ can be Gaussian or Lorentzian) around
the signal frequency  $\delta$, with a characteristic width of
$T_{\varphi}^{-1}$ - one over the coherence time of the signal. For
two frequencies $\delta_{2},\,\delta_{1}$, the line shape is
the sum of these functions $g\left(\left(x-\delta_{1}\right)T_{\varphi}\right)+g\left(\left(x-\delta_{2}\right)T_{\varphi}\right)$.
It can be shown that for a Lorentzian line-shape, the uncertainty diverges with $\left|\left(\delta_{2}-\delta_{1}\right)T_{\varphi}\right|^{-1}$ - this is the Rayleigh
curse. For example, in the case where $\left(\delta_{2}-\delta_{1}\right)T_{\varphi}=0.1$,
the experiment needs to be prolonged by a factor of $10^2$ to get
the same uncertainty as in the $\left(\delta_{2}-\delta_{1}\right)T_{\varphi}\gg1$
regime. For a Gaussian line-shape the scaling is $\left|\left(\delta_{2}-\delta_{1}\right)T_{\varphi}\right|^{-3}$
when $T_{\varphi}$ is unknown (see SI-\ref*{SI-sec:FI of power spectrum}).

In the next section we show that in the case of a phase sensitive
measurement, the divergence is due to the loss of information on the initial phase
of the signal. This information is always lost whenever estimating the frequencies from the power-spectrum/auto-correlation alone, which explains the divergence in this case.

\begin{figure*}
\subfloat[]{\includegraphics[width=0.32\textwidth]{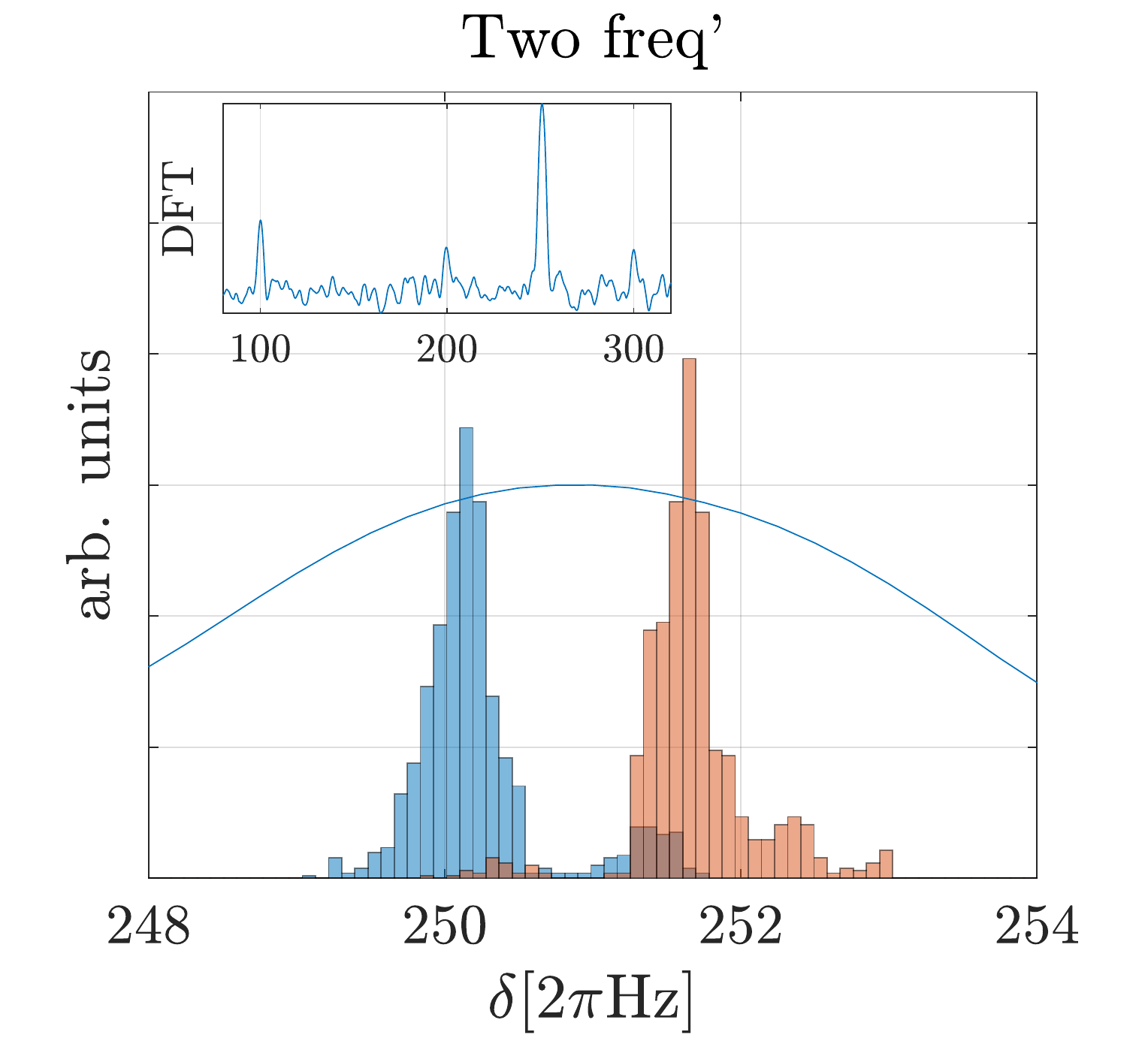} \label{exper_two}}
\hfill
\subfloat[]{\includegraphics[width=0.32\textwidth]{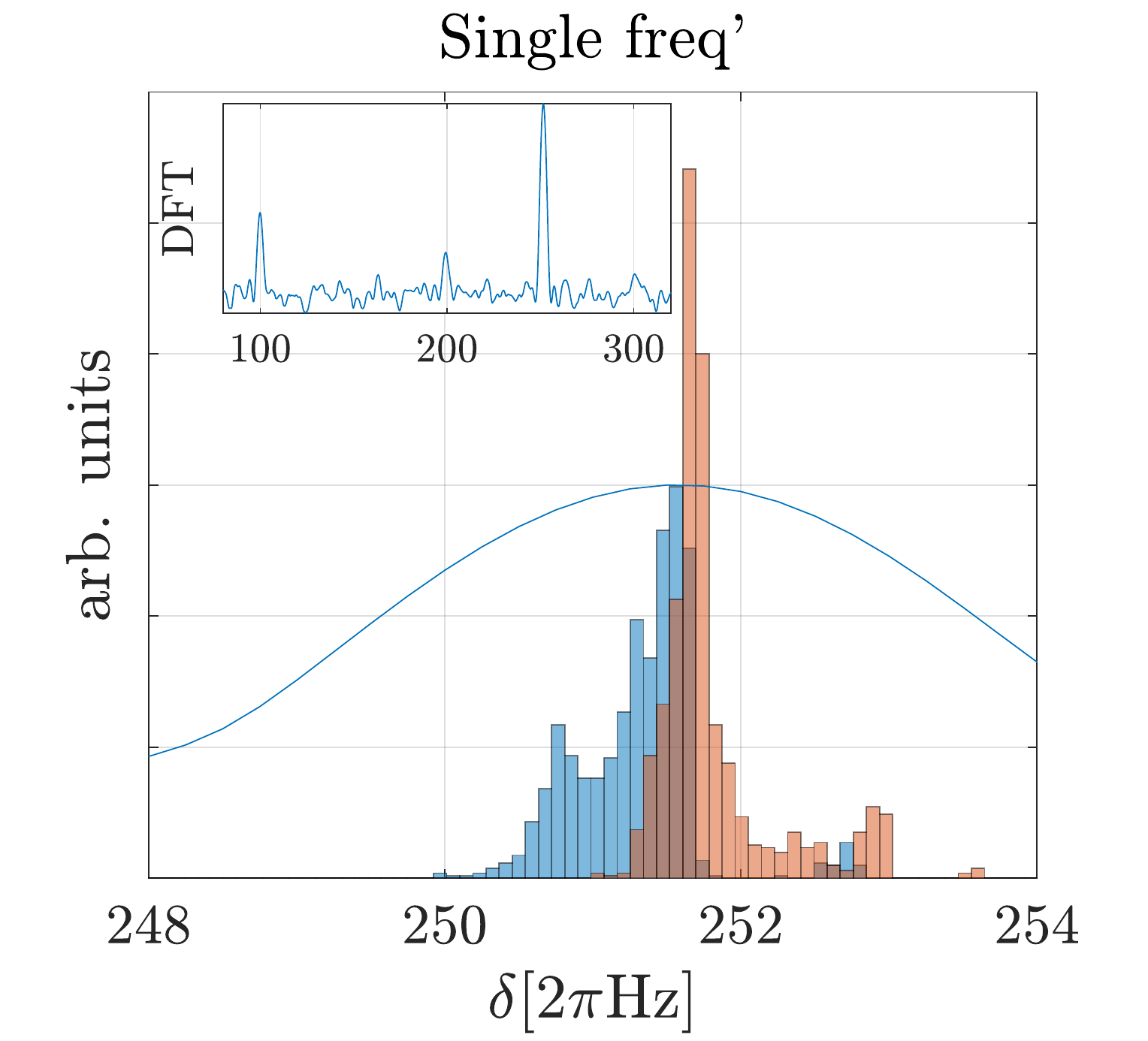} \label{exper_single}}
\hfill
\subfloat[]{\includegraphics[width=0.32\textwidth]{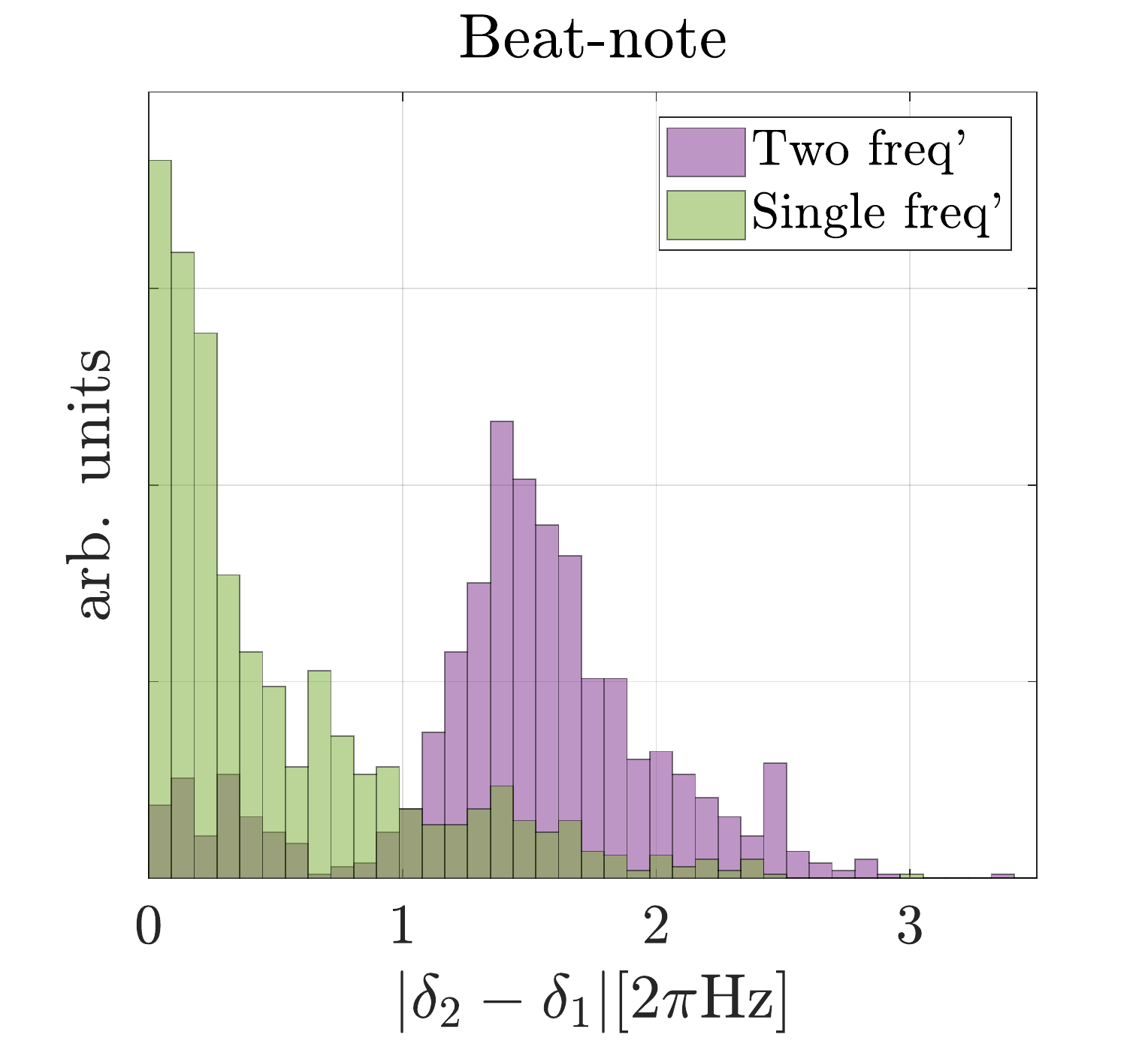} \label{exper_diff}}
\caption{\emph{Resolving the experimental data.} (a) Two frequencies with a frequency difference below the DFT limit, $\delta_1=250\blb{2\pi \text{Hz}}$ and $\delta_2=251.6\blb{2\pi \text{Hz}}$ ($\delta_2-\delta_1=0.4 \blb{2\pi T^{-1}_\varphi}$) were resolved. The Rabi-frequency of the signal was $\Omega\approx 12\blb{2\pi \text{kHz}}$. The inset, and the blue line over the histogram, show the DFT of $33$ measurement sets. The figure depicts a histogram of the estimators from $2^{10}$ iterations of MLE, each over the 33 data-sets randomly chosen from the total of $880$ data-sets. The average estimators are $\protect\blc{\tilde{\delta}_{1}}=250.22\pm0.45\protect\blb{2\pi\text{Hz}}$, and $\protect\blc{\tilde{\delta}_{2}}=251.68\pm0.40\protect\blb{2\pi\text{Hz}}$ (the errors represent the SD); i.e., over $2.4\sigma$ apart. (b) The result of the same procedure, for data containing only a single frequency. The average of the difference is $\protect\blc{|\tilde{\delta}_{1}-\tilde{\delta}_{2}|}=0.51\pm0.53\protect\blb{2\pi\text{Hz}}$. Figure (c) shows the histogram of $|\delta_2-\delta_1|$ taken from the data in figures (a) ({\color{mypurple} purple}) and (b) ({\color{mygreen} green}). Note that the $100\blb{\text{Hz}}$ component in the spectrum (and harmonics) is noise due to fluctuations in the room's fluorescent lights.}
\label{exper} 
\end{figure*}

\paragraph{Phase sensitive measurement of fully correlated signal---}
\label{sec:full correlated signal}Let us start by analyzing a single
frequency signal  under  dynamical decoupling \citep{slichter2013},
where $\delta T_{\varphi}\gg1$. The phase sensitive measurement protocol
is illustrated in Fig.~\ref{scheme_block}, where many measurements
are performed during a phase correlation time and thus the initial
phase can be estimated as well. In the first examples of phase sensitive
measurement on coherent signals, it was shown that the single frequency
estimation uncertainty takes the form of \citep{Qdyne}: 
\begin{equation}
\Delta\bla{\delta}\propto\frac{1}{\Omega\sqrt{\tau}T^{\nicefrac{3}{2}}\sqrt{N}},
\label{FI_limit}
\end{equation}
where $T$ is the minimum between the coherence time of the signal and the stability time of the clock, $\tau$ is the
length of each experimental run and is limited by the coherence time
of the probe ($T_{2}$), $\Omega$ is the Rabi frequency, and $N$
is the number of uncorrelated measurement sets. The prefactor changes according to the precise experimental realization (readout
efficiency, prior knowledge of the amplitudes, etc.). The optimal
SD presented in Eq.~\ref{FI_limit} can be attained
by Maximum-Likelihood Estimation (MLE), which saturates the Cram\'er-Rao
bound \citep{cover2012elements}. The use of MLE has been shown to
be useful in quantum optics in recent years \citep{Brakhane2012,Gammelmark2014,Six2015,Qdyne}.

Eq.~\ref{FI_limit} deals with a single frequency estimation, and since
we are interested in resolution, the quantity of interest is $\Delta\bla{\delta_{1}-\delta_{2}}$
(or $\Delta\bla{\delta_{1}}+\Delta\bla{\delta_{2}},$ in some cases)
in the limit of $|\delta_{1}-\delta_{2}|T_{\varphi}\ll1.$ According
to the Cram\'er-Rao bound: $\Delta\bla{\delta_{j}}\geq\sqrt{\bla{I^{-1}}_{\delta_{j},\delta_{j}}}$,
where $I$ is the Fisher-Information (FI) matrix, and $\bla{I^{-1}}_{x,y}$ is the value relevant for the convenience between $x$ and $y$ taken from the inverse matrix of $I$. Therefore it is critical
to inquire whether $\bla{I^{-1}}_{\delta_{j},\delta_{j}}$ does not diverge
as $|\delta_{1}-\delta_{2}|T_{\varphi}\ll1$, this analysis appears
in SI-\ref*{SI-sec:qdyne_FI}, where we show that under certain
conditions this quantity remains approximately constant. The intuition
for these results can be understood from looking at the single frequency
case where $\delta T_{\varphi}\ll1$ which is analogous to the case
of two close frequencies where we know the average frequency. For dynamics
given by the Hamiltonian $\mathcal{H}=\Omega  \sigma_z \cos\left(\delta t+\varphi\right)$,
the probability for detecting a photon is given by 
\begin{align*}
p & =\sin^{2}\left(\phi+\frac{\pi}{4}\right)\\
\phi & =\Omega\tau\sinc\left(\frac{\delta\tau}{2}\right)\cos\left(\delta t+\varphi\right)
\end{align*}
 where $\phi$ is the phase gained by the probe during the measurement,
and $\Omega,\varphi,\delta$ are the unknown parameters; Rabi frequency, phase, and frequency, respectively. In the limit
of $\Omega\tau\ll1$ and $\delta T_{\varphi}\ll1$ the probability
can be simplified to 
\begin{equation*}
p\approx\frac{1}{2}+\Omega\tau\left(\cos\left(\varphi\right)-\sin\left(\varphi\right)\delta t\right).
\end{equation*}
In this case we can only estimate $\Omega\cos\varphi$ and $\Omega\delta\sin\varphi$ and this is the underlying cause of the resolution problem. In the case where the amplitude is constant, which corresponds to a more classical signal, intuitively we can get some information about $\delta$ by performing more measurements, each with a different phase, and analyzing the statistics.

More precisely, for a single set of measurements (with the same phase) there are three degrees of freedom ($\delta,\Omega,\varphi$) to be estimated, but only two constraints for $\Omega\cos\varphi$ and $\Omega\delta\sin\varphi$. Each new measurement set adds two more constraints, but only one degree of freedom; hence, the frequency can be estimated. Here we see that the phase noise is crucial for the estimation of $\delta$.
In the case where the amplitude also fluctuates, which corresponds to signals commonly seen in the nano-NMR scenario, each new measurement adds two degrees of freedom and the resolution problem re-emerges. In this case, by averaging the probability function over the amplitude distributions the problem effectively reverts back to the model with constant amplitude, and enables us to estimate $\delta$.
In the next sections we show numerically and experimentally that this intuition holds true and indeed the resolution limit is no longer valid. See SI-\ref*{SI-sec:Resolution Limits} for more information.

\begin{figure}
\includegraphics[width=0.99\columnwidth]{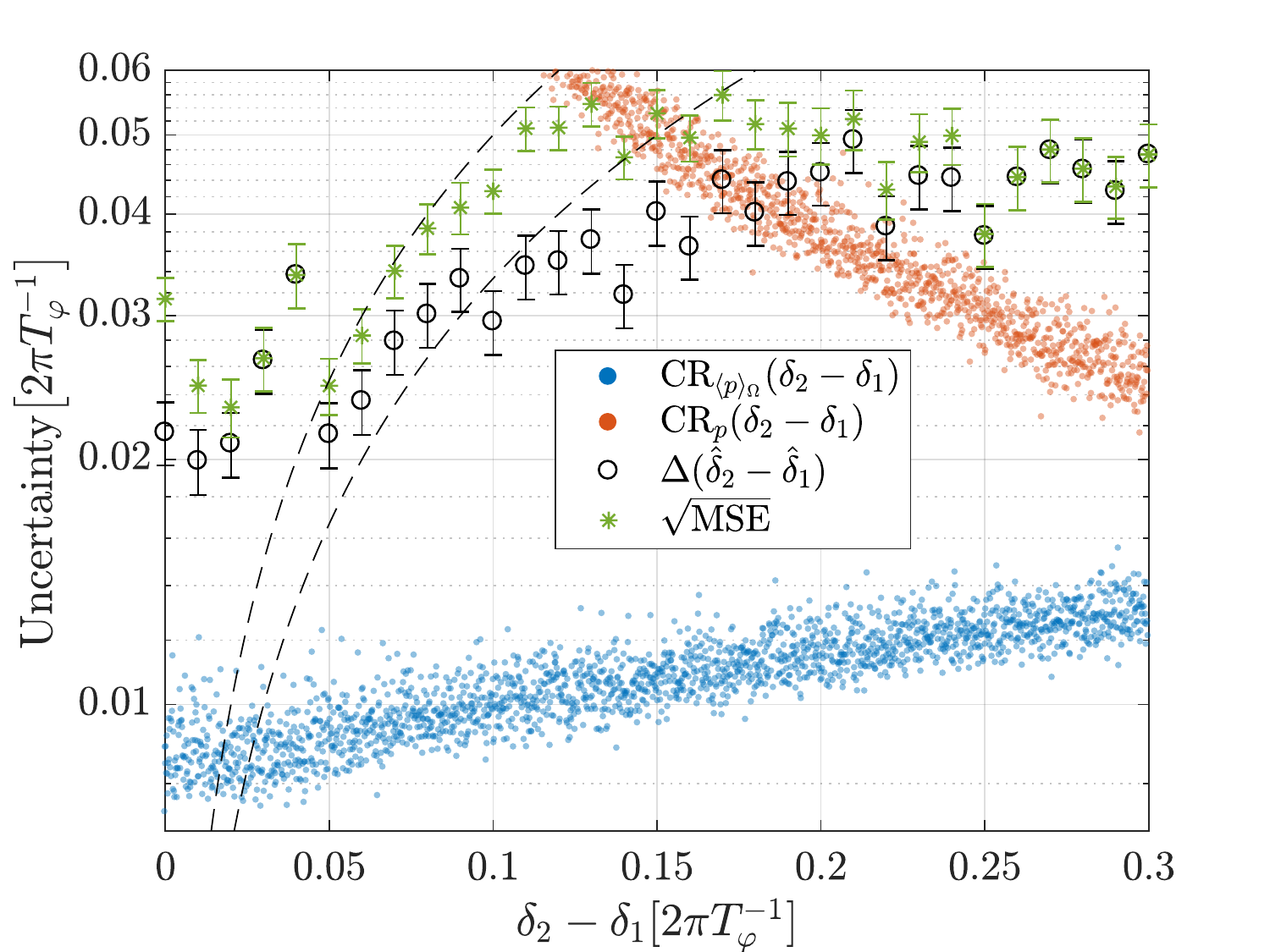}
\caption{\emph{SD of estimators from Global search vs. Cram\'er-Rao bound.} The SD of the estimators are plotted in black circles with  error bars calculated from the SD of the statistics, and the root of the mean square error is plotted in {\color{mygreen} green} asterisks. For comparison the Cram\'er-Rao bound of $\delta_{2}-\delta_{1}$ calculated from the probability ({\color{myred2} red}), and the average probability over the amplitudes ({\color{myblue2} blue}) - the scattering is due to the randomness of the phases. The dashed black line indicates the 2-3$\sigma$ certainty regime. The signal is generated with two frequencies of $\delta_{1}=e^{4}\left[2\pi T_{\varphi}^{-1}\right]$ and $\delta_{2}=\delta_{1}+\Delta\delta$, with phases randomly taken from a uniform distribution $\varphi\sim U\left(0,2\pi\right)$ and amplitudes randomly taken from a Rayleigh distribution $\Omega\sim R\left(\sqrt{2/\pi}\left\langle \Omega\right\rangle \right)$.}
\label{GS for NMR like signal compared with the FI}
\end{figure}

This basically challenges a key assumption in the field of nano NMR
\citep{Qdyne,boss2017,Bucher2017,Pfender2016,Laraoui2013,ajoy2015,Bucher2017,Staudacher2013,muller2014}:
Namely, that resolution is set by the time $T_{\varphi}$ (see Fig.~\ref{scheme_block})
and extra uncorrelated measurements will not greatly improve the ability
to resolve two frequencies. We claim that with phase sensitive measurement,
the resolution behaves like the single frequency uncertainty (Eq.~\ref{FI_limit}).
This is shown by analytic calculation of the FI (SI-\ref*{SI-sec:qdyne_FI})
and numerical simulation of MLE (Fig.~\ref{GS for NMR like signal compared with the FI}) \citep{SI_ref}.

\paragraph{Experimental results---}
\label{sec:experimetal results}

To verify the theoretical analysis we performed high resolution radio-frequency spectroscopy with a single NV center in diamond. The experimental procedure is depicted in \cite{Qdyne} which involved a phase sensitive measurement where a series of resonant $\pi$ pulses were applied to the NV center in the form of an XY8 sequence. The spacing of the $\pi$ pulses provides a spectral filter centered around $1/2\tau=500\blb{\mathrm{kHz}}$, with a band-width dictated by the number of $\pi$ pulses (here $N=8$, which corresponds to a bandwidth of $\sim$60kHz). After application of one XY8 sequence, the NV center is read out to give information about fluctuating magnetic fields within the spectral bandwidth. By concatenating several of these measurements and precisely recording the elapsed time between measurements, the line-width of the resulting spectrum can be further reduced to be limited by $T^{-1}_\varphi$. Here a total coherent measurement time of 0.25 seconds was used to obtain a measurement line-width of 4Hz. These measurements were repeated to generate $880$ uncorrelated measurement sets. For more information see SI section \ref*{SI-Experimental procedure}. This scenario imitates the case in which the resolution is limited by the clock coherence time ( $T_\varphi$).

The ML analysis is shown in Fig.~\ref{exper}. The estimation procedure is as follows. First, a Fourier transform analysis is done to obtain an initial estimate of the amplitude, and the central frequency. The minimization procedure is a multi-search around the initial estimates of the frequencies - with a width of $5T_{\varphi}^{-1}$, and the phases are generated randomly from a uniform distribution. The uncertainty obtained for the frequencies is around $0.4$ Hz; i.e., an order of magnitude smaller than the Discrete Fourier Transform (DFT) limit, which is the line-width of the power spectrum.

\paragraph{Nano NMR signal---}
\label{sec:nano NMR signal}

We now consider the case of the nano NMR signal; i.e., a signal that originates from an unpolarized substance such that both the amplitude and the phase fluctuate. Let us analyze the model that originates in unpolarized nano NMR \cite{Staudacher2013,Aslam,pham,Kong2015}, where the polarization is normally distributed around zero and does some random process; e.g., Ornstein-Uhlenbeck, with a finite coherence time $T_{\varphi}$. We analyze the signal at short times compared to the coherence time; i.e., we assume the signal is totally coherent up to time $T_{\varphi}$, and then the amplitude and phase are redrawn from their respective distribution; e.g., Rayleigh distribution for the amplitude and uniform for the phase. 

Unlike the previous noise model the regular Bayesian analysis results in a diverging frequency uncertainty when the two frequencies merge into one.
In order to overcome this divergence we use a different probability function for the Bayesian fit which we get by averaging over the amplitudes. By doing so we aim to revert to the previous noise model and remove the divergence; see SI-\ref*{SI-sec:random_amplitudes} for more details.

The numerical analysis is shown in Fig.~\ref{GS for NMR like signal compared with the FI}.
The signal is generated with two frequencies of $\delta_{1}=e^{4}\left[2\pi T_{\varphi}^{-1}\right]$ and $\delta_{2}=\delta_{1}+\Delta\delta$, with phases randomly taken from a uniform distribution $\varphi\sim U\left(0,2\pi\right)$ and amplitudes randomly taken from a Rayleigh distribution $\Omega\sim R\left(\sqrt{2/\pi}\left\langle \Omega\right\rangle \right)$. Where the mean amplitudes are $\left\langle \Omega\right\rangle =0.05\left[2\pi\tau^{-1}\right]$, each sequence of measurement has a length of $T_{\varphi}=2^{10}\tau$, and $2^{6}$ sequences with different phases and amplitudes are generated; i.e., each signal is composed of $2^{6}\cdot2^{10}$ measurements, and for simplicity we assume single shot measurements.
The estimation procedure is the same as for the experimental data (Fig.~\ref{exper}) up to the fact that we have used an effective distribution after averaging over the amplitude. The procedure is repeated $2^{7}$ times to gather statistics. Comparing the multi-search procedure with the local search around the real parameters did not always find the most likely estimators, which indicates that with more computing power the global search should converge to the estimators from the local search (see SI Fig.~\ref*{SI-LS for NMR like signal compared with the FI}); furthermore, we believe that as more datasets will be analyzed together the SD of the estimator should converge to the blue dots in the figure. A pronounced difference is shown between the regular Bayesian method (red) which diverges and the corrected Bayesian method (blue), for which the Fisher Information tends to a constant value.

\paragraph{Conclusions and outlook---}
We showed theoretically and verified experimentally that the resolution achieved for spectroscopic measurements by a quantum sensor does not diverge for small beat-note frequencies ($|\delta_2-\delta_1|T_\varphi\ll1$) but rather scales like precision; i.e., as $1/\sqrt{N}$, where $N$ is the number of measurements, with a proportionality coefficient comparable to the single frequency case. This contrasts with the paradigm that the scaling should be much worse, to such an extent that it limits the resolution to the line-width.

We analyzed a general scenario of amplitude and phase noise.
We believe that it captures the essence of a large family of noise models that exhibit temporal fluctuations in phase/amplitude.
In particular, we have studied in detail the exact model of noise that originates from the lack of clock stability and the noise model imposed by diffusion in the nano-NMR setting.
In both cases we have shown that unlike in the regular macroscopic NMR setup, the line-width does not impose a limit on frequency resolution. 

This super-resolution method utilizes phase and amplitude noise in nano-NMR experiments to increase resolution. It is interesting to draw an analogy to optical super-resolution algorithms that utilize noise such as STORM \cite{storm} or SOFI \cite{SOFI1}.


\paragraph{Acknowledgments---}
A. R. acknowledges the support of ERC grant QRES, project No. 770929 and the collaborative project ASTERIQS. 
T.G. acknowledges the support of the Adams Fellowship Program of the Israel Academy of Sciences and Humanities.
J. P. acknowledges MINECO FEDER funds FIS2015-69512-R and Fundaci\'on S\'eneca (Murcia, Spain) Project No. ENE2016-79282-C5-5-R.

\end{document}